\documentclass[twoside]{article}
\usepackage{times}
\usepackage[latin1]{inputenc}
\usepackage{t1enc}
\usepackage{a4}
\usepackage{epsfig}
\usepackage{graphicx}

\def\title{\noindent \Large \bf }
\def\author(s){\large \sc
\vspace*{4mm}\hspace*{11.6mm}\parbox[t]{158mm}}
\def\institute(s){\normalsize \rm
\vspace*{2mm}\hspace*{11.6mm}\parbox[t]{158mm}}
\def\text{\normalsize \rm
\vspace*{5mm}\hspace*{11.6mm}\parbox[t]{158mm}}

\begin{document}




\title{
N-body/SPH study of the evolution of dwarf galaxies in a
cluster environment
}

\author(s){
Dolf Michielsen$^1$, Sven De Rijcke$^1$, Herwig Dejonghe$^1$
}

\institute(s){ 
$^1$Sterrenkundig Observatorium, Ghent University,
Krijgslaan 281, S9, B-9000 Gent, Belgium\\
dolf.michielsen@ugent.be 
}

\text{

Using an N-body/SPH code, we explore the scenario in which a dwarf
elliptical galaxy (dE) is subjected to ram-pressure stripping due to
the intracluster medium (ICM). We suppose that the dwarf galaxy
contains an ionized interstellar medium (ISM) with a temperature
of$10^4$\,K, in hydrostatic equilibrium with the Burkert dark matter
potential. We varied the mass of the dark matter halo from $10^6$ to
$10^{10}$\,M$_\odot$, while the relative velocity of the galaxy
through the cluster was fixed at 1000\,km\,s$^{-1}$. Initially, the
ionized ISM is set up to be pressure confined within the surrounding
ICM. This is in contrast to the simulations of Mori \& Burkert (2000),
where the ISM is density confined. Also, in Mori \& Burkert, the ISM
temperature varies with dark matter mass ($T_{\rm \small ISM}\propto
M_{\rm dm}^{4/7}$), ranging from $6\times10^2$\,K to
$1.3\times10^5$\,K. Since the ISM is supposed to be ionized by
supernova heating, a constant temperature of $10^4$\,K is presumably
more physically motivated.\\

In a typical cluster environment ($\rho_{\rm \small ICM} =
10^{-4}$\,cm$^{-3}$, $T_{\rm \small ICM}=10^{7}$\,K) we find that
smaller dwarf galaxies ($M_{\rm \small dm} < 10^8$\,M$_\odot$) are
instantaneously stripped of their ISM, confirming the results of Mori
\& Burkert. However, the central pressure in the more massive dwarfs
($M_{\rm \small dm} \geq 10^9$\,M$_\odot$) is considerably higher than
in the Mori \& Burkert simulations, due to the lower
temperatures. Therefore, these dEs are able to retain their ISM over
several dynamical time-scales (up to a few Gyrs), during which the gas
may cool and form stars. Star formation is furthermore enhanced by the
increased density due to the ram-pressure of the ICM. In a next step,
radiative cooling, star-formation and stellar feedback will be
included to realistically capture the effects of the compression of
the interstellar medium. Thus, our simulations show that while (i)
smaller dEs lose their ISM almost immediately after entering the
cluster, (ii) more massive dEs are able to retain their gas for
considerable timespans.\\

Although dEs were mostly believed to be gas-poor systems, having lost
their gas either through a galactic wind or by ram-pressure stripping
in a dense group/cluster environment, recent observations show that
some dEs in cluster environments are indeed able to retain some of their gas
and even show evidence for recent or ongoing star formation. (See
e.g. De Rijcke et al., 2003 and Michielsen et al., 2004 for detailed
observations of H$\alpha$ in Fornax dEs.)\\

References:\\

De Rijcke, S., Zeilinger, W. W., Dejonghe, H., Hau, G. K. T., 2003,
MNRAS, 339. 225\\
Michielsen, D., De Rijcke, S., Dejonghe, H.,
Zeilinger, W. W., Prugniel, P., Roberts, S., 2004, MNRAS, submitted
Mori, M., Burkert A., 2000, ApJ, 538, 559
}

\vfill
\end{document}